\documentclass[12pt]{article}
\usepackage{mainstyle}
\usepackage[backend=bibtex, maxnames=4, sorting=none, citestyle=numeric-comp, bibstyle=trad-unsrt]{biblatex}
\usepackage[a4paper,
            left=2cm,
            right=2cm,
            top=2cm,
            bottom=2cm]{geometry}

\usepackage{placeins}
\usepackage{subfig}

\title{Fighting Newtonian Noise with Gradient-Based Optimization at the Einstein Telescope}

\author{Patrick Schillings\footnote{patrick.schillings@rwth-aachen.de}\, and Johannes Erdmann}

\date{\small RWTH Aachen University, III. Physikalisches Institut A, Aachen, Germany}

\begin{document}

\maketitle

\begin{abstract}
\noindent
Newtonian noise in gravitational wave detectors originates from density fluctuations in the vicinity of the interferometer mirrors.
At the Einstein Telescope, this noise source is expected to be dominant for low frequencies.
Its impact is proposed to be reduced with the help of an array of seismometers that will be placed around the interferometer endpoints.
We reformulate and implement the problem of finding the optimal seismometer positions in a differentiable way.
We then explore the use of first-order gradient-based optimization for the design of the seismometer array for 1~Hz and 10~Hz and compare its performance and computational cost to two metaheuristic algorithms.
For 1~Hz, we introduce a constraint term to prevent unphysical optimization results in the gradient-based method.
In general, we find that it is an efficient strategy to initialize the gradient-based optimizer with a fast metaheuristic algorithm.
For a small number of seismometers, this strategy results in approximately the same noise reduction as with the metaheuristics.
For larger numbers of seismometers, gradient-based optimization outperforms the two metaheuristics by a factor of 2.25 for the faster of the two and a factor of 1.4 for the other one, which is significantly outperformed by gradient-based optimization in terms of computational efficiency.
\end{abstract}

\section{Introduction}
\label{sec:Intro}
The observation of gravitational waves (GW) from binary black-hole mergers~\cite{firstdirectGWmeasurement} with the LIGO~\cite{LIGOpaper} and Virgo~\cite{Virgopaper} experiments confirmed the prediction of GWs by Einstein's theory of general relativity and opened up the field of GW astronomy.
Since then, many more GW signals were measured with LIGO and Virgo, including GWs from binary neutron star mergers~\cite{LIGOScientific:2017ync} and mergers of neutron stars and black holes~\cite{LIGOScientific:2021qlt}.
The Einstein Telescope (ET)~\cite{ETDesignReportUpdate} is proposed as a next-generation GW detector.
It is designed to significantly improve over the sensitivity of the LIGO, Virgo and KAGRA~\cite{KAGRApaper,Aso:2013eba} interferometers and aims to measure fainter signals and in particular signals at lower frequencies.

In order to suppress noise from seismic activity, the experiment is proposed to be placed underground and suspension systems 
will decouple the interferometer mirrors from the ground, thus attenuating the seismic displacement. 
Seismic waves traveling through Earth, however, also lead to tiny density fluctuations of the rock surrounding the mirrors.
For low frequencies (below approximately 10~Hz), the resulting gravitational force is indeed a limiting factor for GW measurements~\cite{NewtonianNoiseOrigin,ETsensitivityCurve, EarlyLIGO_NN, VirgoNN, ALowerLimitOnNN}.
The interferometer cannot be shielded from this so-called ``Newtonian noise'' (also sometimes called ``gravity gradient noise'').
However, the surrounding seismic wave field can be measured with seismometers in order to predict the movement of the interferometer mirrors---a technique that has already been successfully demonstrated at the Virgo experiment~\cite{VirgoNewtonianNoiseCancellingSystem}.
Since ET is to be built underground, boreholes need to be drilled for the seismometers and their positions hence need to be chosen with special care~\cite{Amann:2022xyz, FrancescaSingleMirrorOptimization}.

In Ref.~\cite{FrancescaSingleMirrorOptimization}, a Wiener filter was used to predict the Newtonian noise at the mirror positions based on the signals from the surrounding seismometers.
An analytical model was constructed that describes
the noise residual for a single interferometer mirror after subtracting the Wiener filter prediction. 
This model was then used as a cost function in the optimization of the seismometer positions using metaheuristics.
This study was extended to the joint optimization of all mirrors in one interferometer endpoint using the maximum of the residuals as the cost function~\cite{FrancescaJointMirrorOptimization}.
These studies investigated the optimization of the seismometer positions
up to a number of $N=20$~seismometers and 
showed
that the optimal positions depend on the assumed properties of the seismic field, in particular the frequencies that are used for optimization and the fraction of P-waves (primary/pressure waves) and S-waves (secondary/shear waves).

In order to reach optimal seismometer positions in the $3N$-dimensional parameter space, the optimizations in Refs.~\cite{FrancescaSingleMirrorOptimization,FrancescaJointMirrorOptimization} were repeated 100 times, finally picking the trial with the lowest residual.
As a result, the optimization is costly in terms of computing time and hence limited to $N\leq 20$.
In addition, the results from the 100 trials displayed clear symmetries in the seismometer positions that, however, were not reflected in single trials.
While the best single trial may correspond to the global minimum, this might also be an indication that the metaheuristics did not converge to the minimum of the cost function.

In this work, we study the prospects of gradient-based techniques for the optimization of the seismometer positions for Newtonian noise suppression.
We find that only one modification is necessary to reformulate the problem in a differentiable way and we implement the resulting cost function in JAX~\cite{jax2018github} to make use of JAX's support for automatic differentiation.
We then use Adam~\cite{AdamAlgorithm} as an example of a well-established first-order optimization algorithm and compare its performance in terms of the achievable residual and an estimate of the optimization time.
Such efforts to automate the optimization of experimental design with gradient-based methods have been studied in the areas of particle physics~\cite{MODE:2021yid,MODE:2022znx,Dorigo:2020cae,Shirobokov:2020tjt,Neubuser:2021uui,Aehle:2023wwi,Aehle:2024ezu}, muon tomography~\cite{Strong:2023oew} and astrophysics~\cite{Dorigo:2023rbu,Aehle:2023wwi}.

In Section~\ref{sec:FrancExplaination}, we introduce the cost function for Newtonian noise mitigation.
We give details on the implementation of the gradient-based optimization approach in Section~\ref{sec:implementation} and present our results in Section~\ref{sec:results}.
In Section~\ref{sec:conclusions}, we provide our conclusions.

\section{Optimization of Seismometer Positions with Wiener Filters}
\label{sec:FrancExplaination}
With the Einstein Telescope to be built underground, the dominant type of seismic waves are body waves, of which two types exist: P-~and S-waves. P-waves are longitudinal waves that lead to compression and dilation of rock inside Earth and thus directly change the density around the mirrors. Their speed depends on the actual medium but a typical value is
$c_P\approx\SI{6}{\kilo\meter\per\second}$. 
S-waves travel at a lower speed ($c_S\approx\SI{4}{\kilo\meter\per\second}$). They shift rock perpendicular to their direction of propagation and can lead to density fluctuations whenever there are boundaries of materials of different densities that are shifted closer to (or further from) the interferometer mirrors. P-waves, too, can have this influence, but in the direction of their propagation. As the effect of both waves differs, their relative contributions to the total wave field matter, which is given by the parameter $p$:
\begin{equation}
    p=\frac{\EW{\xi_\text{P}\xi_\text{P}}}{\EW{\xi_\text{tot}\xi_\text{tot}}}
\end{equation}
where $\xi_\text{tot}=\xi_\text{P}+\xi_\text{S}$ is the total displacement field of the rock around the Einstein Telescope and $\xi_\text{P}$ is the displacement field of only the P-waves.
$\EW{\cdot}$ denotes the expectation value.
With this definition, P- and S-waves are assumed to be uncorrelated, i.e.,~we assume that there are no sources of seismic waves nearby.

To calculate the gravitational force on an interferometer mirror generated by a known displacement field of seismic waves, only linear operations are necessary (multiplication, integration). Thus, the Wiener filter~\cite{WienerFilter} can be used to predict the mirror acceleration as it is the optimal linear filter
to reduce the quadratic error
for a given frequency $f$
under the assumption of stationary Gaussian noise \cite{Cella}.
Our Wiener filter exploits the displacement recorded by a finite number $N$ of seismometers in all three spacial directions ($x$, $y$ and $z$ - we assume broadband seismometers).
Under the assumptions of isotropic seismic waves, a homogeneous medium and plane waves, the Wiener filter in frequency space can be calculated analytically~\cite{terrestialGravityFluctuations}. Homogeniety also implies that there is no contribution of surface waves. The resulting residual of the actual Newtonian noise on a mirror $i$ and its Wiener filter prediction is given by:
\begin{equation}
    R_i(\omega)=\frac{\EW{\abs{s(\omega)-\Tilde{s}(\omega)}^2}}{\EW{\abs{s(\omega)}^2}}=1-\frac{\Vec{c}_\text{sd}^\dagger\cdot C_\text{dd}^{-1} \cdot \Vec{c}_\text{sd}}{c_\text{ss}}
    \label{eq:residual}
\end{equation}
where $s(\omega)$ is the Fourier transform of the signal (the Newtonian noise at the mirror $i$) at frequency $\omega=2\pi f$, $\Tilde{s}(\omega)$ is the prediction of the signal from the Wiener filter, $C_\text{dd}$ is the cross power spectral density of the data (seismometer displacement measured in 3D), $\Vec{c}_\text{sd}$ is the cross power spectral density of the data vector with the signal, and $c_\text{ss}$ is the power spectral density of the signal.
The seismometer correlation $C_{dd}$ contains an additional term on its diagonal for the self-noise of the seismometers: the constant signal-to-noise ratio SNR that is given by $\text{SNR} = \sqrt{c_\text{ss} / c_\text{nn}}$, where $c_\text{nn}$ is the power spectral density of the seismometer noise. We assume the noise of different seismometers and different channels of the same seismometer to be uncorrelated.
The square root of the residual is an estimate of the remaining effect from Newtonian noise in the GW amplitude.
Varying the positions of the seismometers changes the cross-power spectral densities in Eq.~\eqref{eq:residual}, hence the information that the Wiener filter can use, and consequently the quality of the noise mitigation.
When optimizing the seismometer positions, the goal is to minimize the residual with respect to the $3N$ seismometer coordinates.
So in total, the parameters of the problem consist of the number of seismometers $N$, the frequency $f$ chosen to calculate the residual, the signal-to-noise ratio SNR and the abundance of P-waves $p$. 

The Einstein Telescope is designed to consist of separately optimized interferometers for low- and high-frequency GWs,
where only the test masses for the low-frequency part are expected to be significantly affected by Newtonian noise. 
For close-by test masses, their Newtonian noise is expected to be correlated.
In Ref.~\cite{FrancescaJointMirrorOptimization}, the maximum of the residuals of the individual mirrors from Eq.~\eqref{eq:residual} was chosen as the objective function for optimization:
\begin{equation}
    R=\max_i\bracket{R_i}
    \label{eq:residualCombination}
\end{equation}
While different design options for the Einstein Telescope have been proposed~\cite{ETDesignReportUpdate,Branchesi:2023mws}, here we study the triangle configuration, where each corner comprises four mirrors from the low-frequency interferometer.
We use the same mirror geometry as in Ref.~\cite{FrancescaJointMirrorOptimization}, where the four mirrors are separated by an angle of $60\gr$ and at a distance from the corner of $\SI{64.12}{\meter}$ and $\SI{536.35}{\meter}$, respectively. For the low frequencies of ET, the mirror cavities are much smaller than the seismic wavelengths so that their extent can be neglected.
While numerical values in the optimization would change when repeating the studies for an L-shaped geometry, we expect the conclusions about the optimization methods to remain valid.
All assumptions are summarized in Table~\ref{tab:assumptions}.

\begin{table}[]
    \centering
    \begin{tabular}{l|l}
    No. & Assumption / parameter values \\
         \hline
         1. & No non-linear effects \\
         2. & Isotropic seismic waves\\
         3. & Homogeneous medium (no reflections, no surface waves)\\
         4. & Plane waves\\
         5. & P- and S- waves uncorrelated\\
         6. & Broad-band seismometers (3 channels) with Gaussian uncorrelated noise\\
         7. & Triangle configuration with values mentioned in the text\\
         8. & Extent of the mirror cavity negligible\\
         9. & SNR$=15$, $p=0.2$, $f=1$ Hz or $f=10$ Hz\\
         10. & $c_p=6$ km/s, $c_s=4$ km/s\\
    \end{tabular}    
    \caption{Overview of assumptions (compare with \cite{FrancescaSingleMirrorOptimization,FrancescaJointMirrorOptimization})}
    \label{tab:assumptions}
\end{table}

\section{Optimization Algorithms and Formulation of a Differentiable Loss}
\label{sec:implementation}
Previous studies~\cite{FrancescaSingleMirrorOptimization,FrancescaJointMirrorOptimization,BadaraccoPhD} made use of metaheuristics to optimize the seismometer positions, including basin hopping~\cite{BHAlgorithm}, particle swarm optimization~(PSO)~\cite{PSOAlgorithm} and differential evolution~(DE)~\cite{DEAlgorithm}.
In this work, we compare the gradient-based optimization with Adam to the results from DE, as this optimizer was selected for the most recent optimizations~\cite{FrancescaJointMirrorOptimization}.
In addition, we compare to optimizations with PSO, as we find good performance at lower computational costs compared to DE.
In the following, we provide a brief description of the different optimization algorithms and then discuss the differentiable formulation and implementation of the residual cost function.

For PSO, one chooses a number of swarm particles $n_\text{swarm}$. Each particle $i$ is assigned a position $\vec{x}_i$ in the space of optimization parameters within given boundaries. In our case, this is the $3N$-dimensional space of the coordinates of the $N$ seismometers. The objective function is evaluated at each particle position. Each particle remembers the lowest function value $\Vec{p}_\text{best}$ it has encountered (personal best). In each step $t$, each particle is assigned a velocity $\vec{v}_i$ that is calculated in the following way:
\begin{equation}
    \vec{v}_i(t+1)=\omega\vec{v}_i(t)+r_1 C_c\bracket{\vec{p}_{\text{best},i}(t)-\vec{x}_i(t)}+r_2 C_s \bracket{\vec{g}_{k,\text{best},i}(t)-\vec{x}_i(t)}
\end{equation}
where $r_1$ and $r_2$ are random numbers between 0 and 1 that are re-sampled in every step, and $\vec{g}_{k,\text{best},i}$ is the lowest function value that all of the $k$ nearest neighbors to particle $i$ have encountered up to step $t$.
The tuneable parameters of the PSO algorithm are $\omega$ (``inertia''), $C_c$ (``cognitive parameter''), $C_s$ (``social parameter'') and $k$.
After each step, the particle positions are updated: $\vec{x}_i(t+1)=\vec{x}_i(t)+\vec{v}_i(t+1)$.

The algorithm stops and returns $\vec{g}_{n_\text{swarm},\text{best}}(t_\text{max})$ after a given number of steps $t_\text{max}$ or, with early stopping, after $\vec{g}_{n_\text{swarm},\text{best}}$ has not changed more than a small tolerance value in the last $P$ steps, with the patience parameter $P$.
We use the PSO implementation in \texttt{pyswarms}\footnote{\texttt{pyswarms.single.global\_best.GeneralOptimizerPSO()}}~\cite{pyswarmsPackage}
with $n_\text{swarm}=800$, $C_c=1.5$, $C_s=2$ and $\omega=0.1$ in a ring topology with $k=80$. These values, except for $n_\text{swarm}$, were chosen in a random search that showed overall little dependency on the parameter values for the example of $f=1$~Hz, $N=3$, $\text{SNR}=15$ and~$p=0.2$.
A tolerance of $1\e{-3}$ is used with a patience of 20 steps and a maximum of 1000 steps.

For DE, one also chooses a number of positions $\vec{x}_i$ in the $3N$-dimensional space, called the population size $n_\text{pop}$. Again, the personal best $\vec{p}_{\text{best},i}$ and global best $\vec{g}_{\text{best},i}$ values of the objective function are saved in each step. To calculate new positions, for each position from the population, three distinct other population members $\vec{x}_a$, $\vec{x}_b$ and $\vec{x}_c$ are drawn. The new positions are then calculated as follows:
\begin{equation}
    x_{i,j}'=\begin{cases}
    x_{i,j} & \text{if } r_j<p_\text{Cross}\\
    x_{a,j}+F\cdot\bracket{x_{b,j}-x_{c,j}} & \text{else}
    \end{cases}
\end{equation}
where $j$ enumerates the dimensions of the position vectors and $r_j$ is a distinct random number between 0 and 1. The tuneable parameters of the algorithm are the crossover probability $p_\text{Cross}$, which should be chosen between 0 and 1, and the differential weight $F$, which is typically set between 0 and 2, but can also be randomly generated for each generation of new positions. The new positions $\vec{x}_i'$ are accepted and replace $\vec{x}_i$ if their function value is equal to or smaller than that of the original position; otherwise, $\vec{x}_i$ remains a member of the population. Again, the final result is the global best after $t_\text{max}$ or early stopping.
We use the DE implementation in \texttt{SciPy}\footnote{\texttt{scipy.optimize.differential\_evolution()}}~\cite{scipyPackage} with a population size of 65 per dimension (so $n_\text{pop}=3N\cdot65$),
$p_C=0.75$ and $F$ drawn between 0 and 1.5 in each generation. We use random initialization, a tolerance of $1\e{-3}$ and a maximum number of steps of 4500.

Both algorithms are metaheuristics that search the parameter space within given boundaries, keep testing new positions that are methodically chosen based on the previous positions and, at some point, settle on the lowest function value they have seen in the process. They can efficiently cover large parameter volumes in high-dimensional parameter spaces and do not require any assumptions on the objective function.
In particular, they can also be used for non-differentiable and non-continuous functions.
However, the higher the dimensionality of the parameter space, the lower the coverage of test evaluations for a fixed population size.
In general, PSO and DE are not guaranteed to converge to local or global minima, in particular for high-dimensional optimization problems.

Adam, on the other hand, starts from a single initial position $\vec{x}_0$ that is gradually improved.
As a variant of stochastic gradient descent, it is
frequently used in the training of neural networks, where the mean of the objective function over a small number of training samples is used to calculate the gradients.
In our use case, however, we use Adam to calculate gradients of a fixed cost function without a stochastic component.
It is based on the gradient
$\vec{g}(\vec{x})$ with $g_i(\vec{x}) = \frac{\dd f}{\dd x_i}(\vec{x})$.
In every step, it updates the position according to 
\begin{equation}
    \vec{x}_t=\vec{x}_{t-1}-\alpha \frac{\vec{m}_t}{\sqrt{v_t}+\varepsilon}
\end{equation}
where the parameter $\alpha$ is the ``learning rate'' and $\varepsilon$ is a very small parameter that prevents division by zero. The value $\vec{m}_t$ is a weighted average of the gradients from past steps, resulting in an inertia in the evolution of $\vec{x}$. Similarly, $v_t$ is the weighted squared derivative:
\begin{align}
    \vec{m}_t=\frac{\beta_1}{1-\beta_1^t} \vec{m}_{t-1}+\frac{1-\beta_1}{1-\beta_1^t} \vec{g}(\vec{x}_t)\\
    v_t=\frac{\beta_2}{1-\beta_2^t} v_{t-1}+\frac{1-\beta_2}{1-\beta_2^t} \vec{g}^2(\vec{x}_t)
\end{align}
The parameters $\beta_1$ and $\beta_2$ are tuneable exponential decay rates for the two moment estimates. In each step, $f(\vec{x})$ is saved. As before, the algorithm is stopped after a certain number of steps (or with early stopping) and then returns the lowest function value that it encountered. 

Adam requires $f(\vec{x})$ to be differentiable, as it makes use of its gradients.
Due to its widespread application in deep learning, it is well understood and has proven applicable to a large variety of tasks.
With the automatic differentiation capabilities of JAX, calculating the gradient becomes very efficient, even in high dimensions.
The downside of Adam as a deterministic optimizer is that it may converge to sub-optimal local minima from a large fraction of possible initial positions while the deeper minima may be highly localized.

Carefully choosing the initial position for Adam may hence be important, so we do not only try random initialization, but we also initialize Adam with seismometer positions PSO or DE. This is motivated by the fact that the metaheuristics are efficient in testing the whole parameter space, while Adam uses the objective function's gradient to converge to a nearby local (or global) minimum.

To make use of automatic differentiation in JAX, we have rewritten the code from Ref.~\cite{FrancescaJointMirrorOptimization} in JAX, making use of JAX's just-in-time compilation for computational efficiency.
While this was mostly straightforward to implement, it required one important change compared to previous optimizations: to make the cost function differentiable, we replaced the $\max()$-operation that combines the individual residuals of the different interferometer mirrors in Eq.~\eqref{eq:residualCombination} with a differentiable operation, for which we chose the mean of the four mirror residuals.

We use the implementation of Adam in \texttt{optax}\footnote{\texttt{optax.adam()}}~\cite{optaxPackage} with a learning rate\footnote{The parameters $\vec{x}_i$ are the $x$-, $y$- and $z$-coordinates of the seismometer positions in meters.} of $\SI{1}{\meter}$ and the default values of $\beta_1=0.9$, $\beta_2=0.999$ and $\varepsilon=\SI{1e-8}{\per\meter}$.
Early stopping with a tolerance of $1\e{-3}$ and a patience of 500 steps is used, while the maximum number of steps is 10000.

In a first step, the optimization is performed with DE, PSO and Adam using random initialization.
Separate optimizations are done for the number of seismometers, $N$, varying between 1 and 70 in steps of 1.
In Refs.~\cite{FrancescaSingleMirrorOptimization,FrancescaJointMirrorOptimization}, the optimization with DE was repeated 100 times, picking the best optimization run.
To achieve comparable results to these previous optimizations, we repeat the optimization with PSO 100 times with different initial values, also picking the best optimization run.
We choose PSO instead of DE, as it is significantly faster than DE, and we limit the repeated runs of PSO for $N>20$ to $N=25$ as well as $N=30$ to $N=70$ in steps of 10.
In a second step, optimization with Adam is performed using the results of the single PSO and DE runs as initialization, as well as---for a few selected values of $N$---using the best optimization from the 100 runs of PSO as initialization.
We study the optimal positions for frequencies of 10~Hz and 1~Hz, which are example frequencies at the very upper and lower boundaries at which Newtonian noise is expected to be relevant for the Einstein Telescope.
We assume an SNR of 15 and a fraction of P-waves, $p$, of 0.2, which is a benchmark that was thoroughly studied previously~\cite{FrancescaJointMirrorOptimization}.

\section{Results}
\label{sec:results}

\begin{figure}[p]
    \vspace{-48pt}
    \centering
    \includegraphics[width=.65\linewidth]{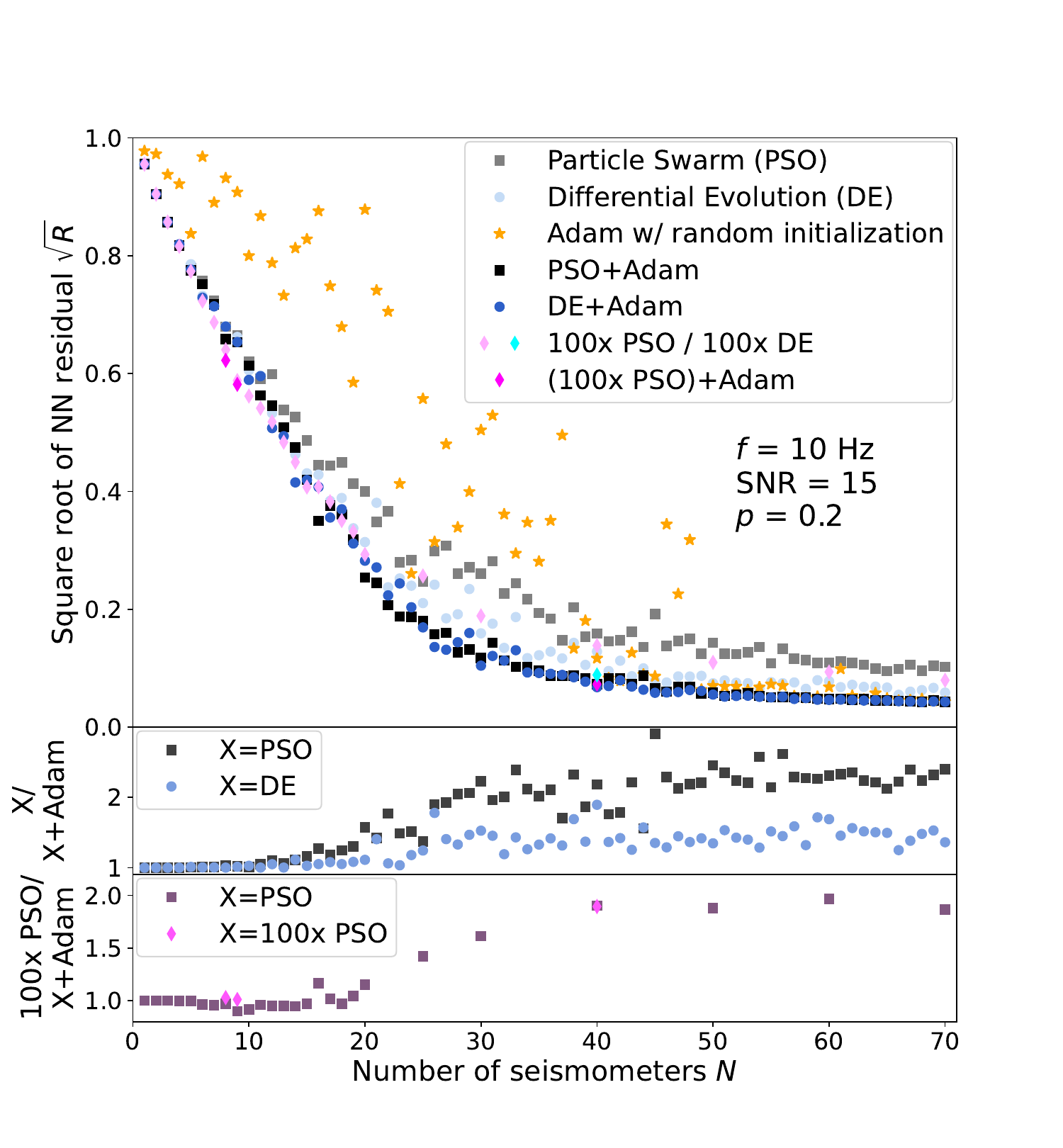}
    \caption{
    $\sqrt{R}$ for Newtonian noise mitigation as a function of $N$ for $f=10$~Hz, $p=0.2$ and $\text{SNR}=15$. The optimization was done with PSO, DE, Adam with random initialization and Adam initialized with PSO and DE, respectively.
    Also shown is the minimum value for 100 distinct runs of PSO (for $N>20$, only $N=25,30,40,50,60,70$ were studied) with a few of them also used as initial values for optimization with Adam. For $N=40$, we also tested the minimum of 100 DE runs.
    The bottom panels show the ratio of the improvement in $\sqrt{R}$ obtained when initializing Adam with single runs of PSO or DE and with a single PSO run or 100 distinct runs of PSO, respectively.
    }
    \label{fig:results}
    \includegraphics[width=.65\linewidth]{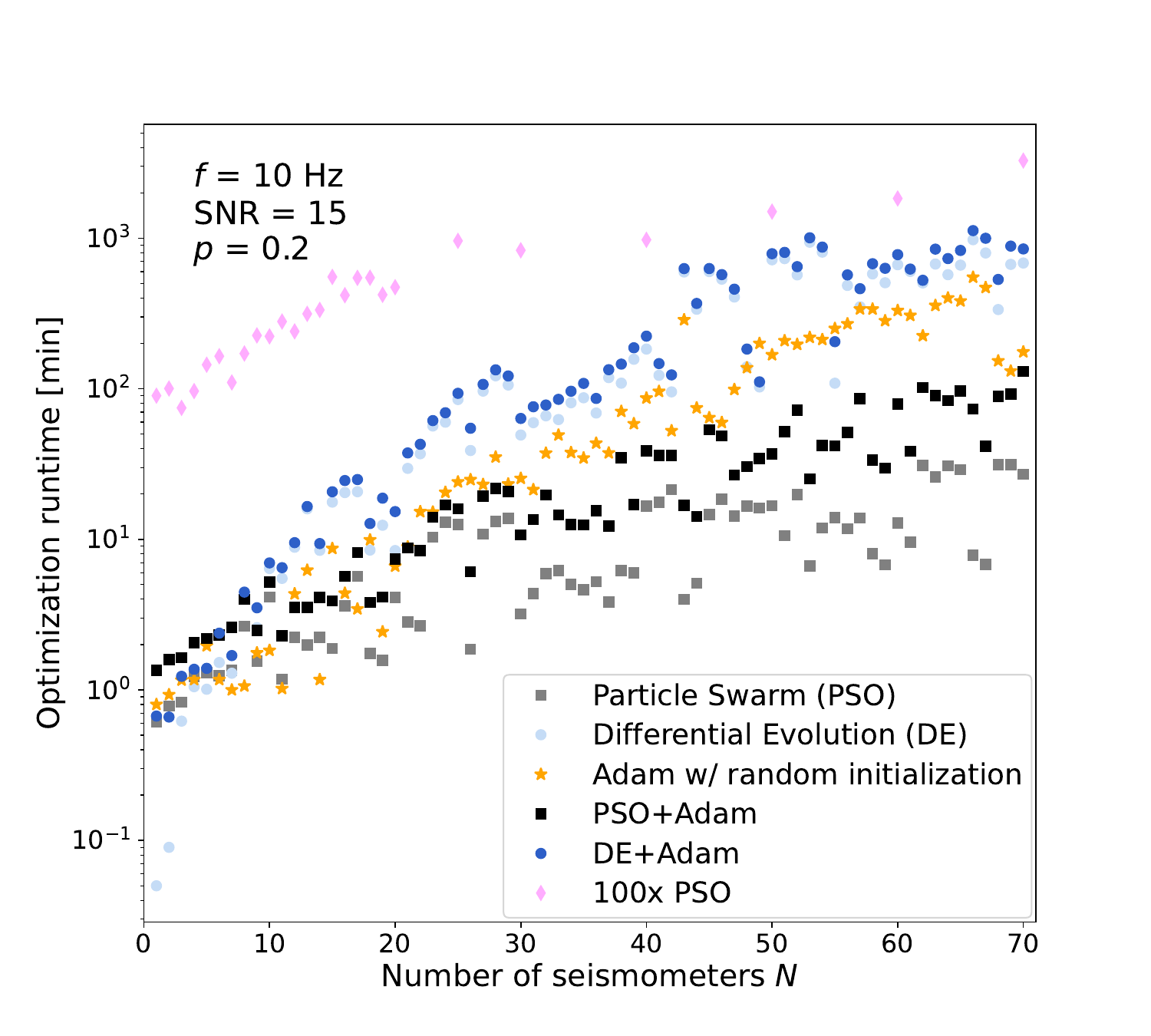}
    \caption{Time consumed for the optimization on the computing cluster by PSO, DE and Adam with random initialization as well as Adam initialized by single runs of PSO and DE and the sum of 100 PSO runs.
    }
    \label{fig:time}
\end{figure}

The results of the optimization at a frequency of $f=10$~Hz are shown in Figure~\ref{fig:results}.
The parameter boundaries for PSO and DE are chosen as $\pm\SI{1000}{\meter}$ in $x$- and $y$-direction and $\pm\SI{300}{\meter}$ in $z$-direction, with the origin of the coordinate system in the interferometer corner.
Shown is the square root of the residual as a function of the number of seismometers $N$. 
In previous optimizations~\cite{FrancescaSingleMirrorOptimization,FrancescaJointMirrorOptimization}, the number of seismometers was limited to $N\leq 20$, motivated by the cost of drilling one borehole per seismometer.
We study also larger values of $N$ in order to quantify the achievable gains in noise mitigation, as future studies may investigate the potential for optimizing the positions of seismometers that can be more cheaply deployed, such as in the interferometer tunnels.

All optimizations in Figure~\ref{fig:results} show the same general trend: they start at a residual close to 1, that is reduced when seismometers are added.
The more seismometers get deployed, the more information can be used by the Wiener filter to predict the Newtonian noise at the interferometer mirrors, to separate the effects of P- and S-waves, and to reduce the impact of the self-noise of the seismometers. The information gain per seismometer is highest when only a small number of seismometers is already deployed.
Random initialization and the intrinsic randomness of the metaheuristics result in fluctuations around this general trend.
These fluctuations can be reduced, for example, by repeated runs of the algorithm, comparing the single PSO run with the best of 100 PSO runs.
Adam with random initialization shows particularly large fluctuations, as its outcome is solely determined by the initial seismometer coordinates.

Comparing single runs of the three optimization algorithms with random initialization, the two heuristic algorithms yield similar results for a small $N$ but deviate around $N=20$ with DE reaching ultimately better residuals for the chosen values of the hyperparameters.
We note that we did not perform a separate hyperparameter optimization for every value of $N$.
In particular, the number of swarm particles in PSO is constant for all values of $N$, while the population size in DE increases with $N$.
The results from Adam with random initialization fluctuate strongly and are generally worse than the PSO and DE results up to $N\approx 50$, with only a few exceptions.
For larger values of $N$, Adam with random initialization often outperforms PSO and DE, which indicates that the exact initialization for large $N$ is not as crucial as for lower values.

Using the results of PSO and DE as initial positions for Adam has no effect for very small values of $N$. This means that in low dimensions, the metaheuristics are able to find values that are close to the bottom of a local minimum (or maybe even global minimum).
Between $N\approx 5$ and $N\approx 15$, the metaheuristics find good values.
Running Adam initialized with these results leads to a slight improvement, but it is often still trapped in the local minimum that was found with the corresponding metaheuristic.
For larger values of $N$, the initialization of Adam with the metaheuristics leads to significant improvements in the optimization. The optimization leads to seismometer positions that are sometimes quite far away from the positions that the metaheuristics had found, resulting in average improvements of factors of 2.25 for PSO and of 1.4 for DE (cf.~the upper ratio plot in Figure~\ref{fig:results}). 
The results after optimization with Adam only differ by a few percent between initialization with PSO or DE, with PSO occasionally yielding the better initialization for Adam.
This means that it is not very important which metaheuristic is used to provide the initial values for Adam.

In Figure~\ref{fig:results}, we also show the results of 100 independent optimization runs of PSO, i.e., with the same hyperparameters but different initial values, choosing the best of all runs as the optimized values. This is comparable to the method that was used in Ref.~\cite{FrancescaJointMirrorOptimization} but with PSO instead of DE.
As expected, this leads to fewer fluctuations in the optimized results.
Again, for small $N$, no significant improvement is visible because already single runs of PSO or DE yield good solutions. Between $N\approx 5$ and $N\approx 15$, this method outperforms all of the single runs of the different optimizers.
For $N=8$ and $N=9$, we tested if there is still potential for improvement by using Adam initialized with this best solution of these 100 PSO runs. Indeed, we find a few percent of improvement (cf. the lower ratio plot in Figure~\ref{fig:results}).
For values of $N$ between 15 and 20, running PSO 100 times yields comparable results to PSO+Adam, but for larger values of $N$, it is outperformed by the single runs of Adam initialized by PSO or DE by a factor of almost~2.
Further optimizing the best of the 100 PSO runs with Adam does not lead to better results than running Adam initialized with a single PSO run, as tested for $N=40$.
Here, we also checked if running DE 100 times can lead to better results, as a single DE run yields lower residuals than a single PSO run, but we found that it is still outperformed by the subsequent runs of a metaheuristic and Adam.

In Figure~\ref{fig:time}, we show the time that was required to calculate the points shown in Figure~\ref{fig:results} on our institute's computing cluster.
The times that we recorded for each run may vary due to the performance of each cluster node and by how many jobs run in parallel on each node, as is visible by the fluctuations of the run times for each optimization strategy.
However, we verified that the general trends and statements in Figure~\ref{fig:time} hold by performing a second independent full optimization run.

As expected, the computing time generally increases with $N$, i.e., with the number of parameters of the optimization task, but the time for the different methods grows differently fast.
With our choice of hyperparameters, PSO runs quickest for high $N$. DE takes about 20 times longer than PSO but also yields better results.
When Adam is initialized with either PSO or DE, it runs much faster than with random initialization, as the criteria for early stopping are met faster given the good choice of initial parameter values.
Even when combining the times of PSO and Adam, it is still below the time for an optimization with Adam with random initialization as well as lower than DE, while giving better residuals. Running PSO 100 times may yield slightly better results at moderate $N$, but can take a factor of 100 longer than the combined optimization of PSO and Adam for these numbers of seismometers. 

\begin{figure}[p]
\makebox[\textwidth]{
    \centering    \subfloat{\includegraphics[width=.5\linewidth]{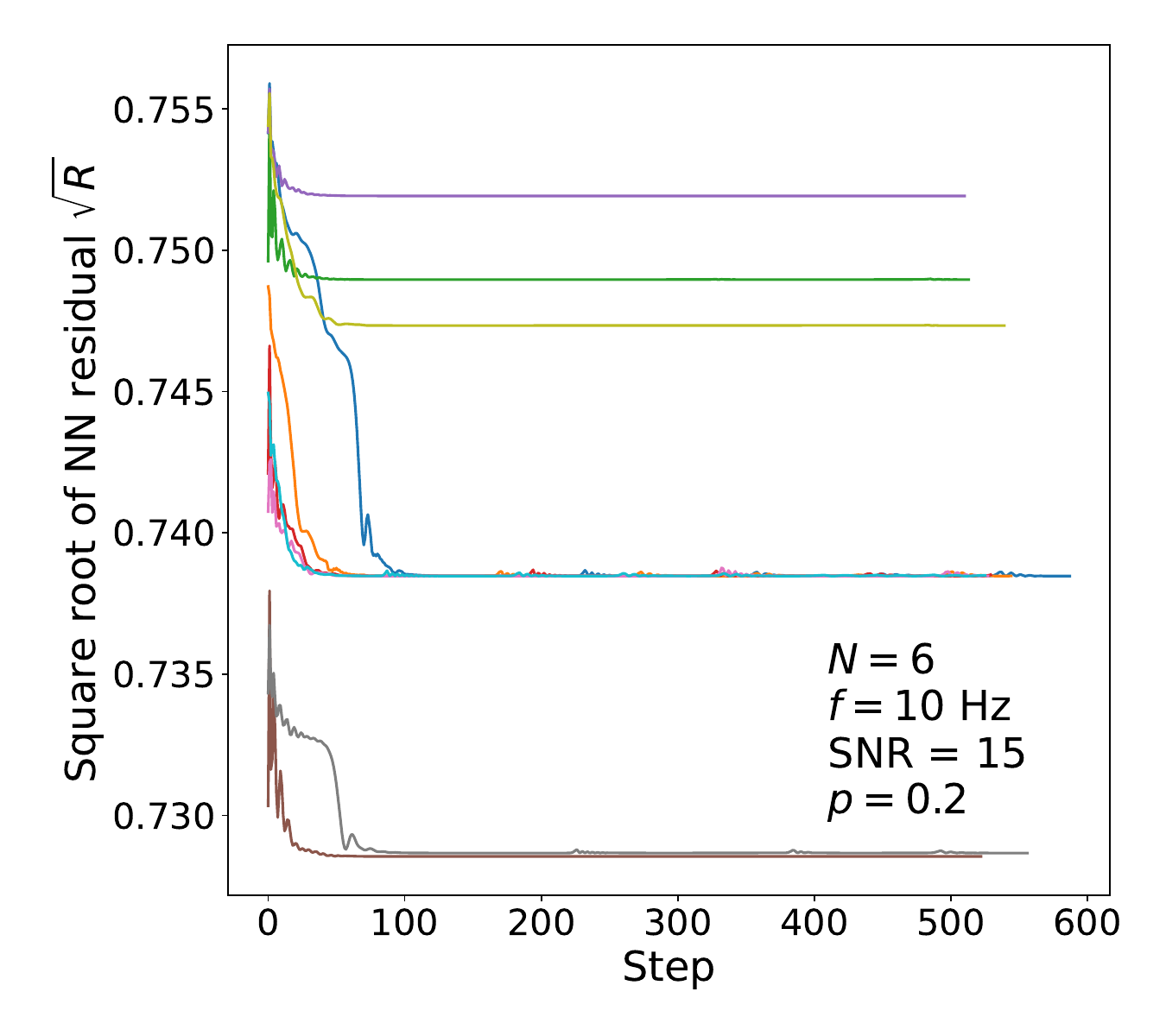}}
    \subfloat{\includegraphics[width=.5\linewidth]{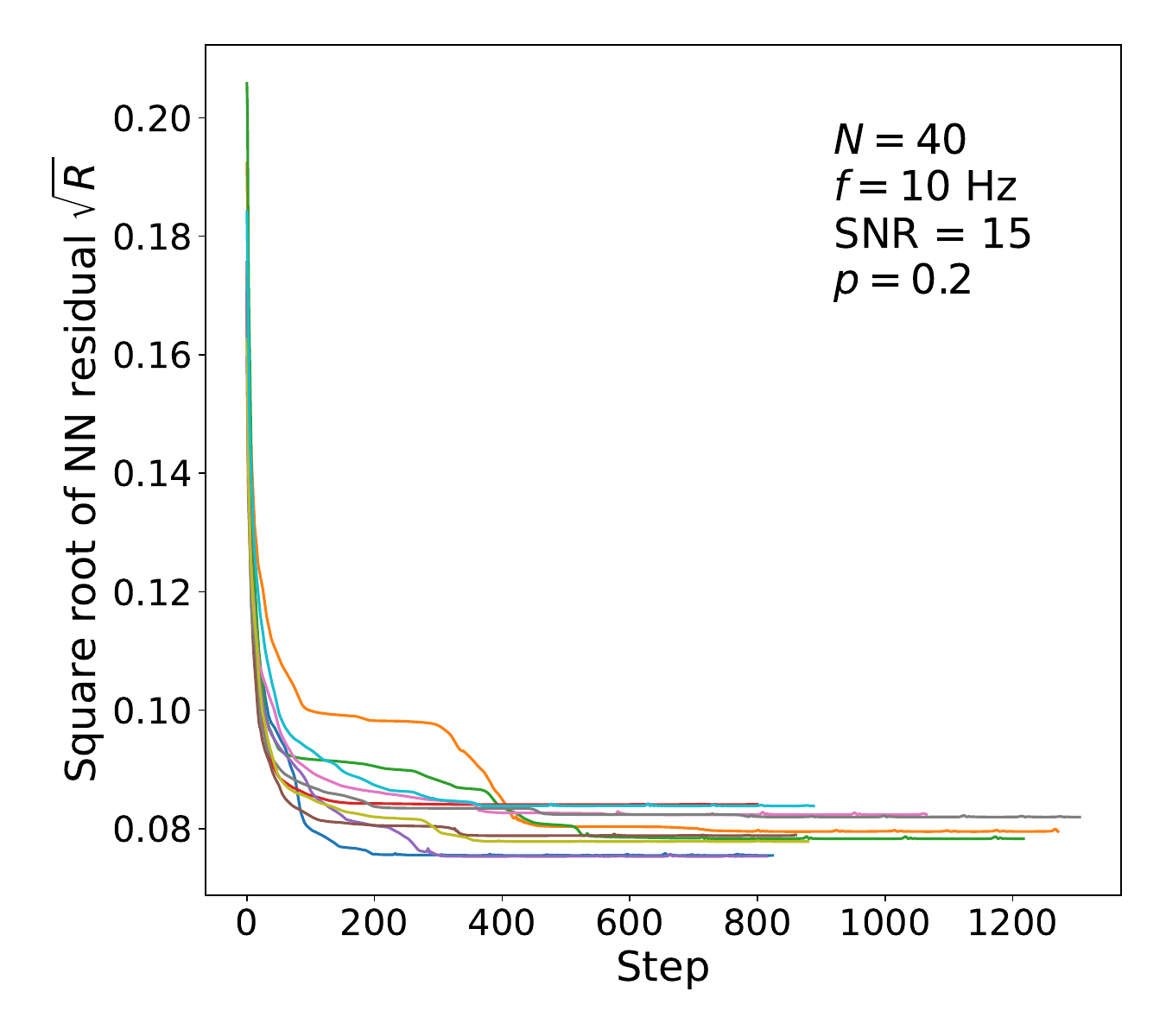}}
    }
    \caption{Evolution of the residual during optimization with Adam for $N=6$~(left) and $N=40$~(right). The initial positions were pre-optimized with PSO. All curves have parameters $f=10$~Hz, SNR = 15 and $p=0.2$ and represent the variation from different PSO solutions.}
    \label{fig:loss}
    \centering
    \includegraphics[width=.7\linewidth]{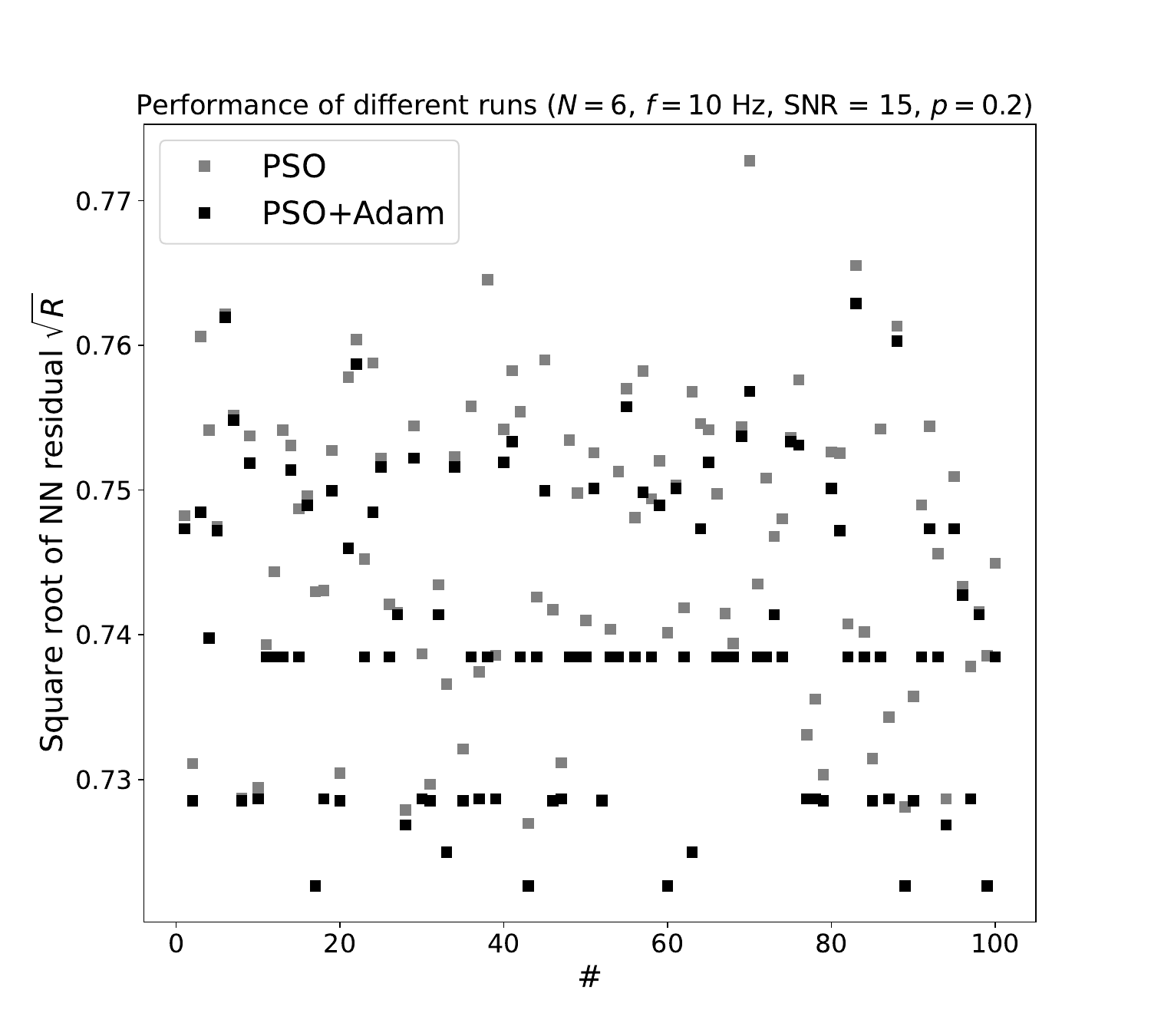}
    \caption{Optimizations with PSO from 100 different runs for $N=6$, $f=10$~Hz, $p=0.2$. The results are shown before (lighter color) and after (darker color) using the PSO results as initial values for Adam.}
    \label{fig:100times}
\end{figure}

The evolution of the residual during the optimization with Adam is shown in Figure~\ref{fig:loss} for $N=6$ and $N=40$ for the initialization with several distinct PSO runs.
The optimization procedure stops after the curve is flat for a while due to early stopping.
For $N=6$, one can see that Adam improves the PSO solutions by about a percent. It is striking that the final solution often depends on the initial positions from PSO and follows the gradient to the bottom of a local minimum.
This can be the same local minimum (or at least one with the same value) for different PSO pre-optimizations.
For $N=40$, the optimization with Adam provides significant improvements for Newtonian noise mitigation and quickly brings the different PSO solutions down to half of the initial residuals.
The spread of the final solutions is small compared to the obtained improvements, and for such large values of $N$, it is hence not guaranteed that a better PSO initial residual will result in a better optimization with Adam.

Figure~\ref{fig:loss} suggests that Adam descends into distinct local minima from the initial PSO results.
We further investigate this for $N=6$ and ask whether these local minima correspond to specific geometries of the seismometer positions.
Residuals of 100 PSO runs before and after further optimization with Adam are shown in Figure~\ref{fig:100times}.
Considering only the results after optimizing with PSO (light squares), they mainly are in the range between $\sqrt{R}\approx0.73$ and $\sqrt{R}\approx0.76$. After running Adam initialized with these PSO results (dark squares), the resulting residuals are not only lower on average but also many optimization runs result in very similar residuals.
We interpret these values as local minima that Adam descends into from the original PSO positions. We also note that the amount of improvement that can be obtained from the optimization with Adam in general cannot be predicted from the PSO result alone.

\begin{figure}[p]
    \centering
    \subfloat[Optimized positions in $yz$-plane (PSO only)] {\includegraphics[height=.52\linewidth]{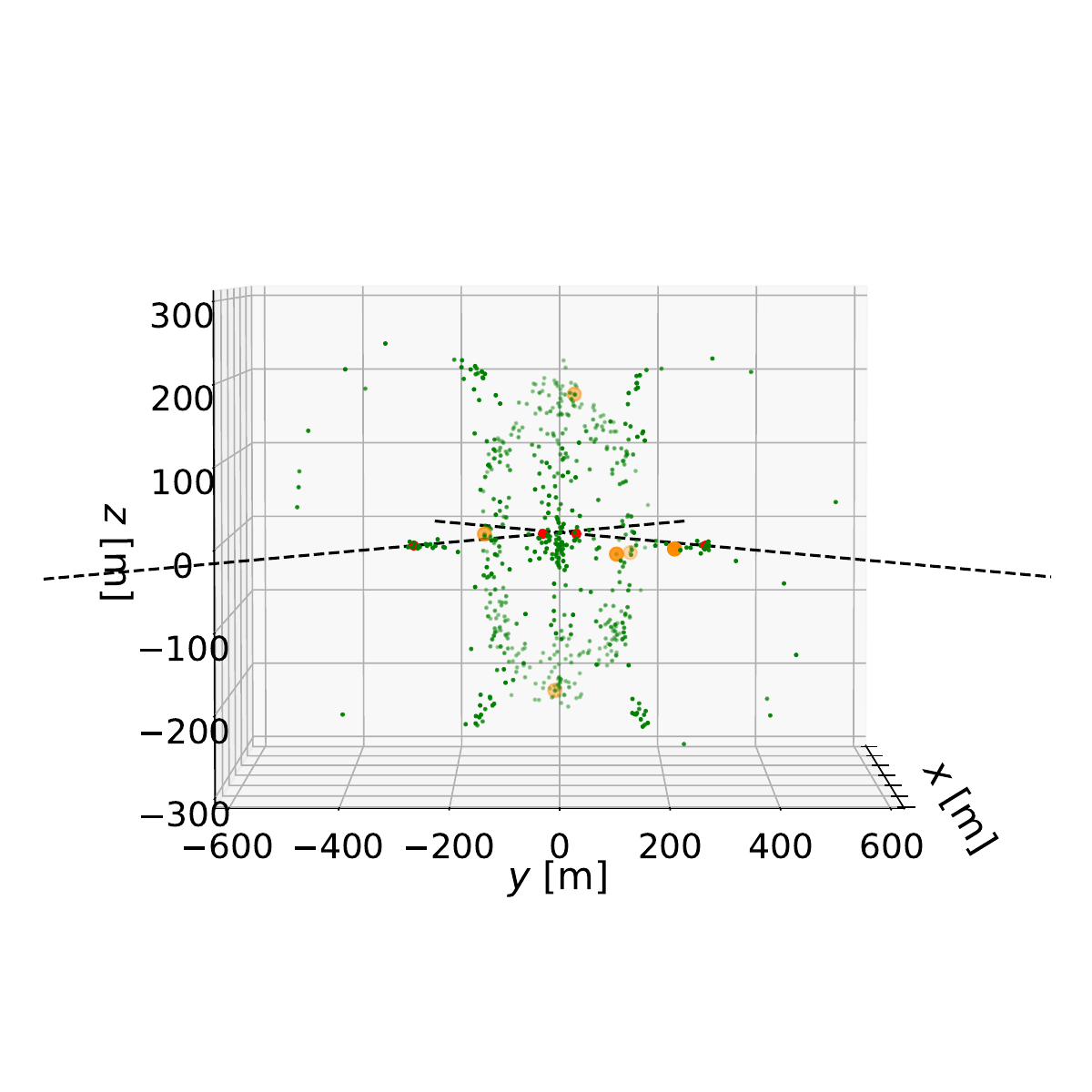}}
    \subfloat[Optimized positions in $yz$-plane (PSO+Adam, $\sqrt{R}\approx0.7384$)]{\includegraphics[height=.52\linewidth]{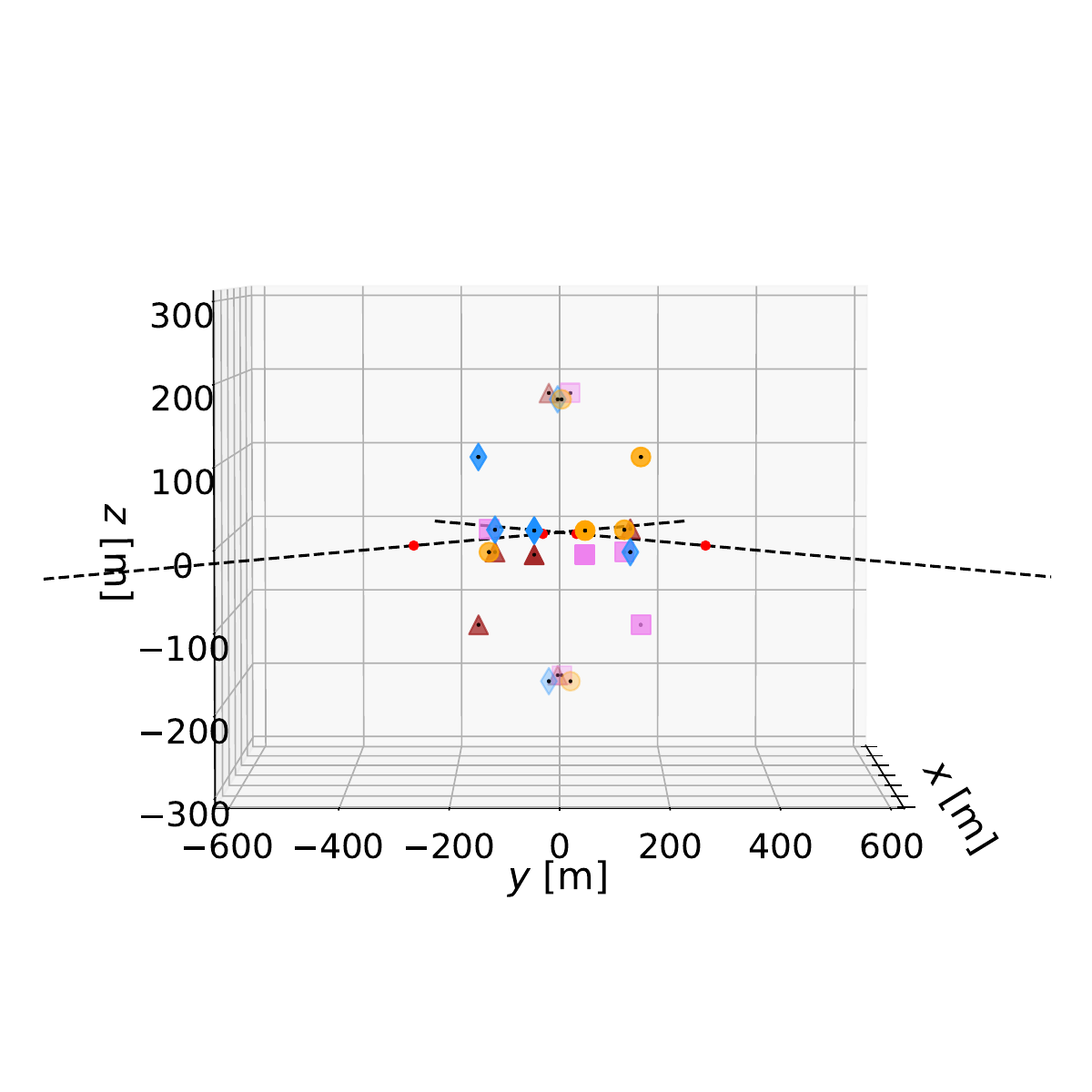}}\\
    \subfloat[Optimized positions in $xy$-plane (PSO only)]{\includegraphics[height=.58\linewidth]{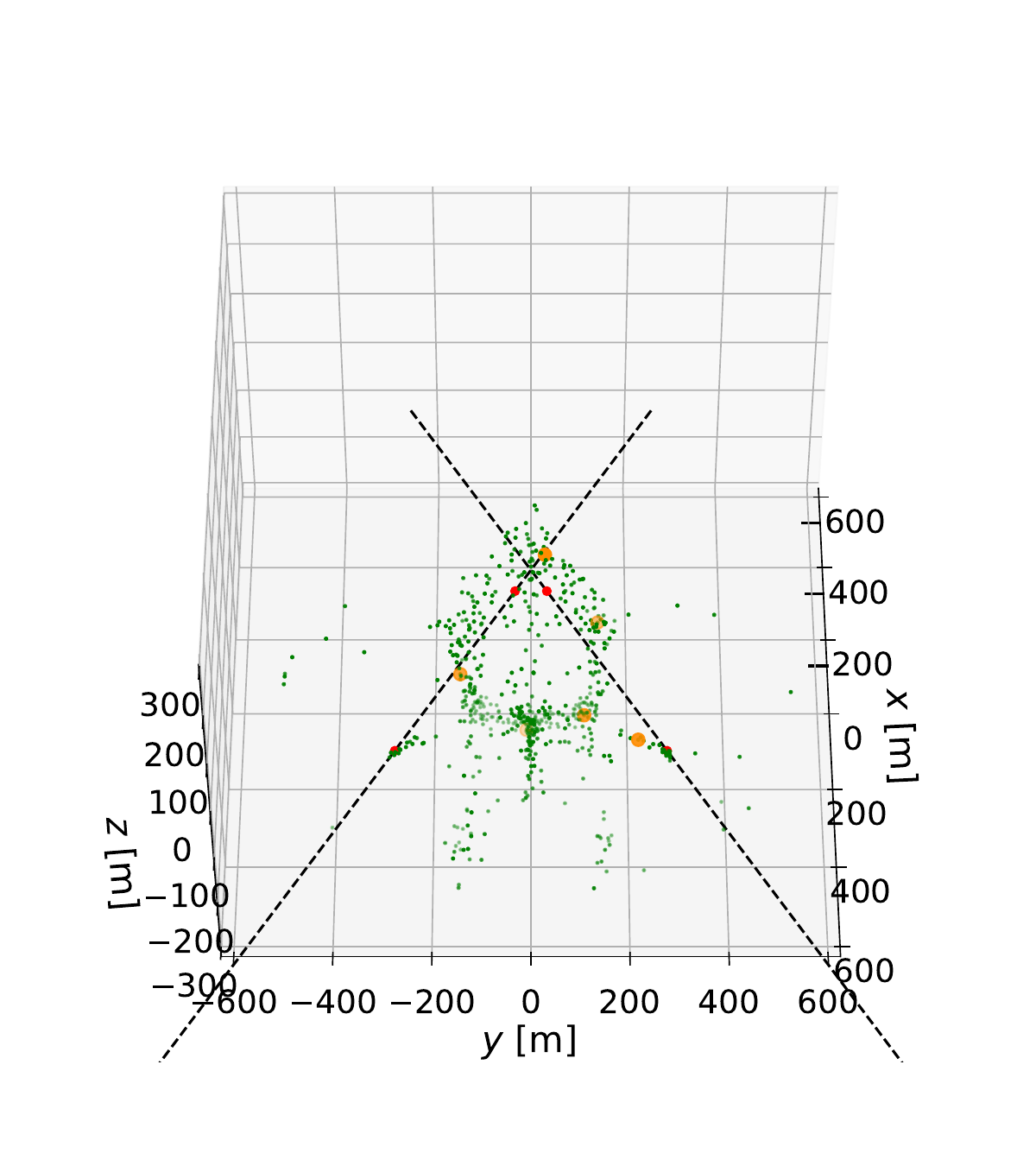}}
    \subfloat[Optimized positions in $xy$-plane (PSO+Adam, $\sqrt{R}\approx0.7384$)]{\includegraphics[height=.58\linewidth]{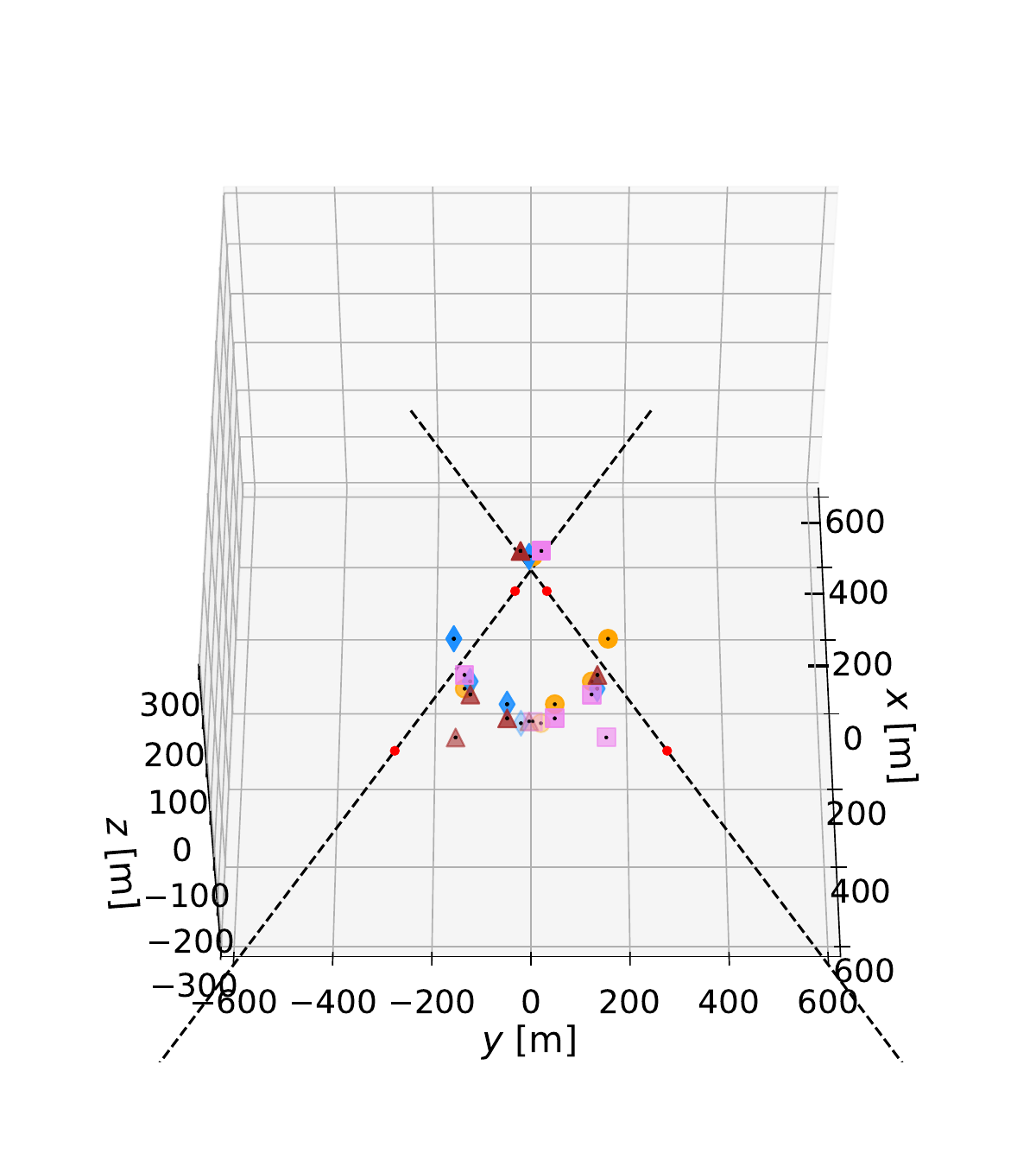}}
    \caption{Resulting positions of seismometers in the $yz$-plane, (a)~and~(b), and in the $xy$-plane, (c)~and~(d), after optimization for $N=6$, $f=10$~Hz, SNR = 15 and $p=0.2$. The red dots and black dashed lines represent the mirrors and tunnels of the Einstein Telescope's corner.
    In~(a) and~(c), all 100 resulting 6 seismometer positions after optimization with PSO only (light points of Figure~\ref{fig:100times}) are drawn together to visualize the symmetry of the setup. Orange points are the positions from the run with the best residual. In~(b) and~(d), all runs that have $\sqrt{R}\approx0.7384$ after optimization with Adam.
    Differently colored symbols show the four different configurations that lead to the same residual, illustrating the symmetry of the task.}
    \label{fig:localMinimumGeometry}
\end{figure}

Figure~\ref{fig:localMinimumGeometry} shows the optimized sensor positions of repeated runs of PSO and PSO+Adam in the $xz$- and $xy$-planes for $N=6$.
In Figure~\ref{fig:localMinimumGeometry}~(a) and~(c), the 6 positions from the 100 distinct runs of PSO (light squares from Figure~\ref{fig:100times}) are shown.
As discussed in Ref.~\cite{FrancescaJointMirrorOptimization}, this shows the symmetries of the task\footnote{We note that the optimization in Ref.~\cite{FrancescaJointMirrorOptimization} was performed with the maximum of the residuals among the four mirrors in one corner and not for their mean.}.
While a symmetry is visible in the ensemble of the optimized positions, which corresponds to the symmetry of the geometry of the task, the run with the lowest residual among all 100 runs (orange points) does not show this overall symmetry itself.

In Figure~\ref{fig:localMinimumGeometry}~(b) and~(d), the optimized positions are shown after PSO+Adam, this time only for runs that result in $\sqrt{R}\approx0.7384$, which represent the runs from the prominent line of dark squares in Figure~\ref{fig:100times}. We identified four distinct solutions among these runs and used different colors and symbols to mark the six seismometers that belong to each solution. One can see that all solutions can be obtained from a single one by mirroring along the symmetry planes of the problem, namely the $z=0$ and $y=0$ planes in this coordinate system.
We interpret this as Adam being able to descend to the local minima of the cost function better than PSO, so that its optimized results very clearly show the symmetries of the optimization task, while the PSO results are smeared around the best positions because they do not descend all the way down to the bottom of the minimum.

While $f=10$~Hz is at the upper end of the relevant frequency range for Newtonian noise, we also ask how Newtonian noise mitigation with the Wiener filter performs for low frequencies and pick $f=1$~Hz as a very low example frequency.
When we optimize for $f=1$~Hz, we find that Adam converges to seismometer positions that are much further away from the mirrors than for $f=10$~Hz.
This is expected because the wavelength for 1~Hz and $c_P\approx\SI{6}{\kilo\meter\per\second}$ ($c_S\approx\SI{4}{\kilo\meter\per\second}$) is as large as 6~km (4~km) for P-waves (S-waves).
Since the analytical model for Newtonian noise mitigation based on the Wiener filter assumes that the rock is isotropically distributed around the mirrors, this model fails to take into account that the Einstein Telescope is expected to be built approximately 300~m underground, so that the optimized seismometer positions must be restricted to a maximum distance in $z$-direction from the mirrors.
While PSO and DE require that a search volume is defined for the optimization, this is not the case for Adam.

To prevent this, we study two methods.
One method is to simply clip Adam's optimization results to the corresponding constraint by letting it finish its optimization and in the resulting positions, replacing every $z$-coordinate that has $\abs{z}>300$~m with $z=\pm \SI{300}{\meter}$.\footnote{It is known that the rock is not homogeneous along the full depth of 300~m, i.e.~the cross correlations of seismometer signals with the Newtonian noise at the mirrors are expected to vary as a function of seismometer depth.
However, for simplicity, we assume that the rock is homogeneous up to $z=300$~m and study how to effectively confine optimization results with Adam to a certain phase space. The same procedure can then be applied to smaller maximum values of $z$, where the rock is still homogeneous to a good approximation.} The other method is to multiply a differentiable function by the residual, which results in a penalty for larger values of $z$.
We propose the function 
\begin{equation}
    R'=R\cdot e^{a\cdot\sum_{i=1}^N\bracket{z_i/600}^b}
    \label{eq:modifiedResidual}
\end{equation}
with $b$ even, as it exponentially punishes $z$-values that are too large, while for $\abs{z}<300$ $R'$ is close to 1. 
Because of the sum in the exponent, if $b$ is small, configurations with large $N$ will be punished more, while for $a=2^b$ and $b\rightarrow \infty$ the modified residual is independent of the number or position of any seismometer inside $z=\pm300$.

\begin{figure}[htb]
    \centering
    \includegraphics[width=\linewidth]{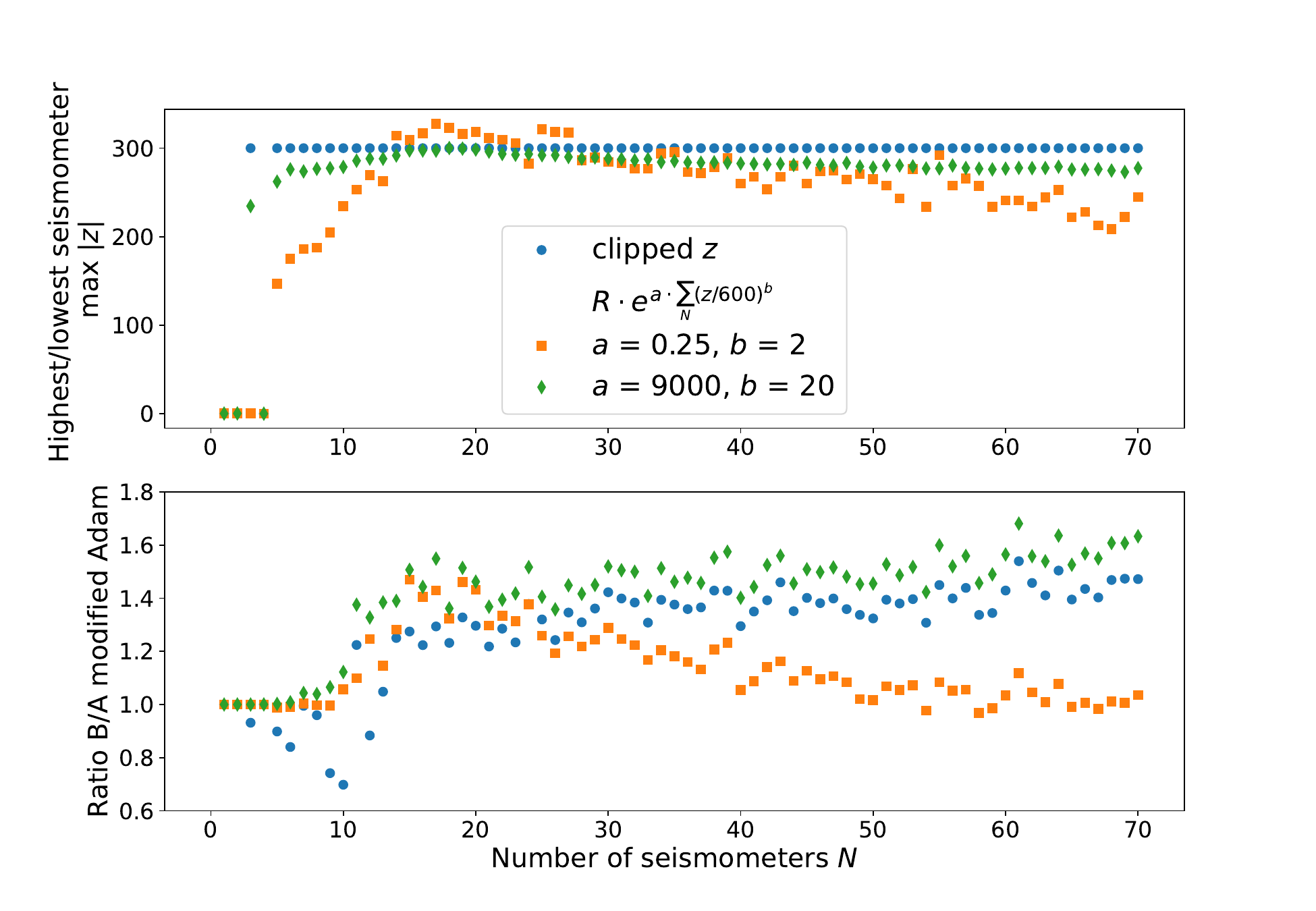}
    \caption{Top: The maximum absolute value of the $z$-coordinates among all seismometers for different modified optimizations with Adam for $f=1$~Hz. The initial positions were chosen with PSO. Bottom: The improvement in the residual after modified optimization with Adam with respect to the PSO-only results,
    which are shown later in Figure ~\ref{fig:f1residuals}.
    }
    \label{fig:modified1Hz}
\end{figure}

For both methods, the residuals are evaluated with the original residual estimate, i.e.,~Eq.~\eqref{eq:residual}, but using the updated $z$-coordinate.
The boundaries for PSO and DE are increased to $\pm\SI{5}{\kilo\meter}$ for the $x$- and $y$-coordinate to account for the larger wavelength.
The boundaries for the $z$-coordinate are kept unchanged. These are then used as initial values for the optimization with Adam, either with the clipped $z$-coordinate or with the constrained cost function from Eq.~\eqref{eq:modifiedResidual}. The resulting $z$-positions for PSO+Adam and the ratio between the PSO-only residual and the residuals for PSO+Adam for the two methods are shown in Figure~\ref{fig:modified1Hz}. 

\begin{figure}[htb]
    \centering
    \includegraphics[height=.7\linewidth]{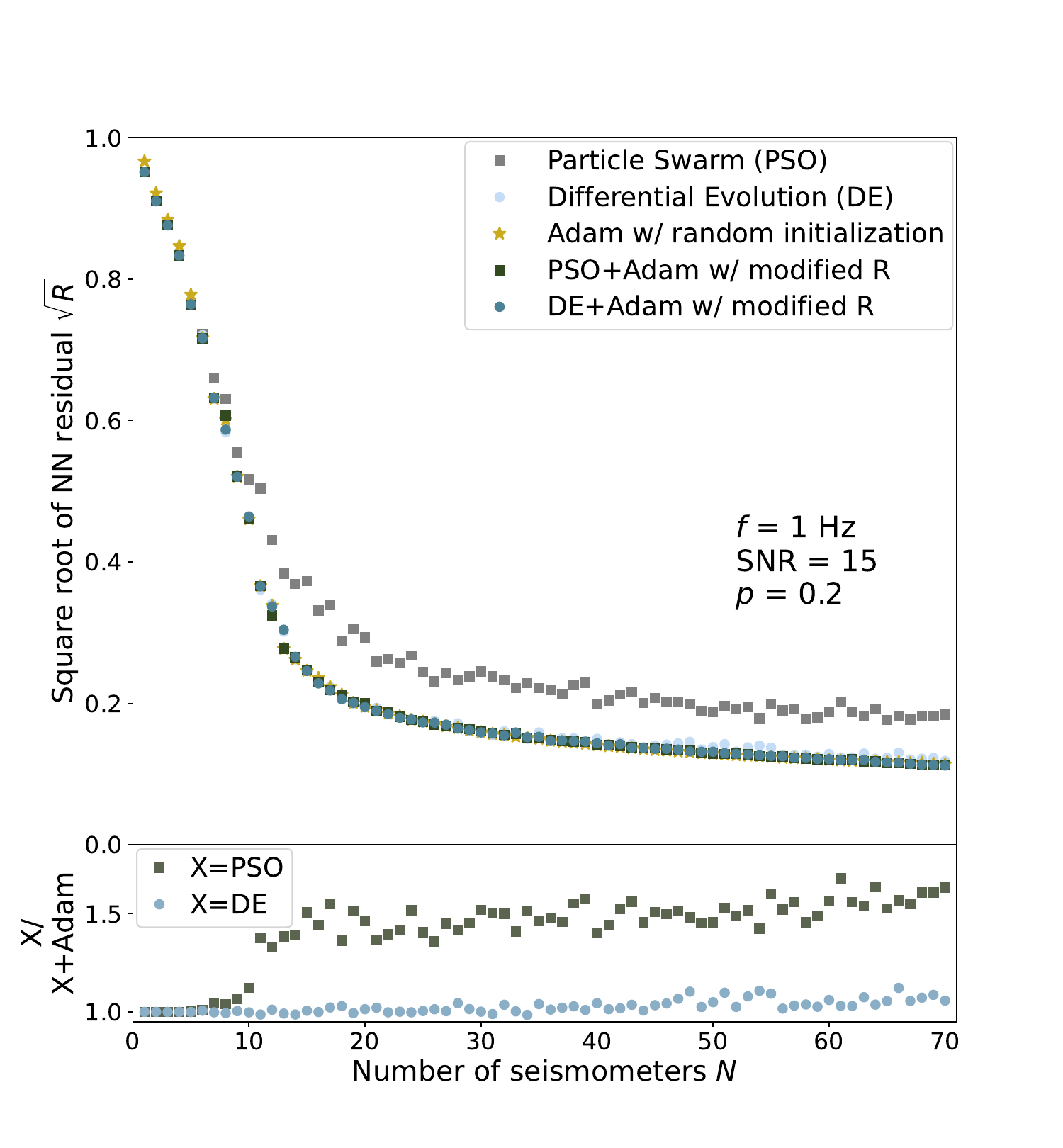}
     \caption{$\sqrt{R}$ for Newtonian noise mitigation as a function of $N$ for $f=1$~Hz, $p=0.2$ and $\text{SNR}=15$.
    The optimization was done with PSO, DE, Adam with random initialization and Adam initialized with PSO and DE, respectively.
    The optimizations with Adam used the modified cost function from Eq.~\eqref{eq:modifiedResidual} with $a=9000$ and $b=20$.
    The bottom panel shows improvement in $\sqrt{R}$ obtained when initializing Adam with PSO or DE compared to optimization with PSO or DE alone.
    }
    \label{fig:f1residuals}
\end{figure}

With both methods, we are able to keep the $z$-positions of the seismometers bounded (upper plot in Figure~\ref{fig:f1residuals}).
By construction, clipping leads a flat line at $z=300$ with a few exceptions at low $N$, where the optimal seismometer positions are in the plane of ET.
The resulting $z$-coordinates for the modified cost function from Eq.~\eqref{eq:modifiedResidual} depend on the number of seismometers and the chosen parameters $a$ and $b$. The curves rise at first up to a maximum at around $N=15$ and then fall for increasing $N$. While $a$ is used to steer the maximum allowed value of the $z$-coordinate, increasing $b$ has the effect of reducing the slope and, therefore, the dependence on $N$.
While for a small number of seismometers, it is not important that the distances of the seismometers from the mirrors in the $z$-direction are large, for larger values of $N$, several seismometers would have optimal $z$-positions of $\geq 300$~m, which results in a larger penalty with increasing $N$ in Eq.~\eqref{eq:modifiedResidual}.

Although it is a simple method, PSO+Adam with clipping improves for large values of $N$ compared to the PSO-only optimization result (lower plot in Figure~\ref{fig:modified1Hz}).
However, this does not work well for some lower values of $N$.
This is not surprising, as the projection of the optimal positions onto $z=300$~m may result in suboptimal $x$- and $y$-positions.
For the second method, the improvement depends on the chosen parameters of the function, $a$~and~$b$.
In Figure~\ref{fig:modified1Hz}, we show results for two extreme choices of $b$, i.e.,~$b=2$ and $b=20$, where the value of $a$ is adjusted to exploit as much as possible from the allowed optimization parameter space.
The resulting values of $a$ are $a=0.25$ and $a=9000$, respectively.
The second choice of parameters works better in terms of constraining the seismometers to $|z|<300$~m and in terms of the achievable residuals.
With these values, we were able to reach an improvement of approximately a factor 1.6 for $f=1$~Hz compared to pure PSO.

Figure~\ref{fig:f1residuals} shows the optimization results for $f=1$~Hz for Adam with the modified cost function with $a=9000$ and $b=20$ for the different initializations (random, PSO, DE) and compares them to optimizations with PSO and DE alone.
Generally, the residuals are larger than for $10$~Hz, as the boundary in $z$-direction hampers the mitigation of Newtonian noise from seismic waves with such large wavelengths.
Otherwise, we observe the same trends as before, with the exception of Adam with random initialization, which performs much better than at 10~Hz and finds good positions for values $N>5$ that are at the same level as the PSO- and DE-initialized optimizations.
The reason is probably that the structures in the optimization space are larger due to the larger wavelength, so that random initialization has a much higher probability to find a good minimum.
While single optimization runs with DE provide very good residuals already, they can still be improved by approximately 5\% with either of the modified Adam optimizations (initialized randomly or with PSO or DE), while Adam initialized randomly or with a single PSO run is faster than single runs with DE by at least the same amount as found for 10~Hz (cf.~Figure~\ref{fig:time}).

\FloatBarrier

\section{Conclusions}
\label{sec:conclusions}
We studied the use of gradient-based techniques for optimizing the positions of seismometers at the Einstein Telescope.
The seismometers are auxiliary sensors that help to reduce the impact of Newtonian noise at low gravitational wave frequencies.
We reformulated an analytical model for the noise residual based on a Wiener filter in a differentiable way and implemented the model in JAX to make use of its automatic differentiation capabilities.
We then optimized the positions with Adam and compared the results to optimizations with differential evolution and particle swarm optimization.
These two metaheuristics had been used in previous studies and were shown to be particularly successful in finding good seismometer positions when repeating the optimization with 100 different initial values.
We studied the optimization as a function of the number of seismometers, where we compared the expected noise reduction and the computational costs of the optimization.
We chose two reference frequencies, 10~Hz and 1~Hz.
For the optimization at 1~Hz, i.e., larger wavelengths, we introduced a constraint term in the optimization to exclude the run-away of seismometer positions outside of the physical volume of the rock.

In general, we find that gradient-based optimization can improve over the optimization with the metaheuristics when initialized with a single optimization run of the metaheuristics.
The achievable gains depend on the complexity of the optimization problem.
In comparison to the single optimization runs, for $f=10$~Hz, the achievable improvements are small up to a number of approximately 10 and 25 seismometers for particle swarm optimization and differential evolution, respectively.
For larger numbers of seismometers, however, improvements of approximately 125\% and 40\% can be achieved at moderate additional computational costs, reaching very similar performance when either initialized with differential evolution or particle swarm optimization.
Better optima can be found by running particle swarm optimization 100 times with different initial values with the associated larger running time.
For large numbers of seismometers, however, the gradient-based optimization still significantly improves over the repeated runs of particle swarm optimization in terms of achievable noise reduction while requiring approximately 50 times less computing time compared to the repeated runs of particle swarm optimization.
In addition, we find for the example of an optimization of six seismometer positions that equally good different solutions from the gradient-based optimization reveal the symmetries of the optimization problem.
For $f=1$~Hz, differential evolution as a metaheuristic is able to find seismometer positions that result in only marginally worse residuals than optimizations with Adam, but at a significantly larger computational cost.
For this frequency, also randomly initialized optimizations with Adam result in a very good mitigation of Newtonian noise.

In this work, we showed the advantages of gradient-based methods for the optimization of the experimental design of a gravitational wave experiment using the example of the seismometer positions at the Einstein Telescope.
While the prospects should be further studied for this particular use case, more applications in the area of gravitational wave experiments may be explored.
For the optimization of the seismometer positions, future directions of research include the exploration of other optimizers, in particular second-order optimizers, and the optimization of the positions under more realistic circumstances.
These include the use of realistic noise spectra from the drilling campaigns at the candidate sites for the Einstein Telescope, more realistic assumptions on the direction of the seismic waves, on the fraction of P- and S-waves, and on models of the rock.
As a fully realistic simulation of the seismic activity at the sites may not be achievable \cite{FrancescaSingleMirrorOptimization}, optimizations that are robust with respect to uncertain parts of the seismic noise model are needed as well.

\section*{Availability of Code}
The code used to produce the results in this paper is available at \url{https://github.com/lc316353/Fighting-Newtonian-Noise-with-Gradient-Based-Methods/tree/main}. 

\section*{Acknowledgments}
The authors thank Francesca Badaracco and Jan Harms for helpful discussions and Francesca Badaracco for providing the original code from Refs.~\cite{FrancescaSingleMirrorOptimization,FrancescaJointMirrorOptimization} and her comments on the manuscript.
This research was supported by the German Federal Ministry of Education and Research (BMBF) via project 05A2023 under grant number 05A23PA1.

\printbibliography

\end{document}